\documentclass[12pt]{article}
\usepackage{amsbsy}
\usepackage{amsfonts}
\usepackage{amssymb}

\usepackage{bm}

\usepackage{enumerate}
\usepackage{graphics}
\usepackage{epsfig}
\usepackage{color}
\usepackage{nicefrac}
\usepackage{mathrsfs}
\usepackage{bbm}
\usepackage{amsmath}

\newlength{\TZ}
\newcommand{\BEQ}{\begin{equation}}     
\newcommand{\BEA}{\begin{eqnarray}}
\newcommand{\BD}{\begin{displaymath}}
\newcommand{\EEQ}{\end{equation}}       
\newcommand{\EEA}{\end{eqnarray}}
\newcommand{\ED}{\end{displaymath}}
\newcommand{\bb}{\begin{eqnarray}}
\newcommand{\ee}{\end{eqnarray}}

\newcommand{\e}{{\rm e}}

\newcommand{\vep}{\varepsilon}          
\newcommand{\D}{{\rm d}}                
 
\newcommand{\demi}{\frac{1}{2}}         

\renewcommand{\vec}[1]{\boldsymbol{#1}} 

\catcode`\@=11
\def\numberbysection{\@addtoreset{equation}{section}
        \def\theequation{\thesection.\arabic{equation}}}
\numberbysection

\begin{document}

\begin{titlepage}

\vskip 1.5 cm
\begin{center}
{\Large \bf On integral representations and asymptotics of some hypergeometric functions in two variables}
\end{center}

\vskip 2.0 cm
\centerline{{\bf Sascha Wald}$^{a,b,}$\footnote{swald@sissa.it} and {\bf Malte Henkel}$^{a,c,d,}$\footnote{malte.henkel@univ-lorraine.fr}}
\vskip 0.5 cm
\begin{center}
$^a$ Groupe de Physique Statistique, 
D\'epartement de Physique de la Mati\`ere et des Mat\'eriaux, 
Institut Jean Lamour (CNRS UMR 7198),  Universit\'e de Lorraine Nancy, 
B.P. 70239, \\ F -- 54506 Vand{\oe}uvre l\`es Nancy Cedex, France
\\ \vspace{0.5cm}
$^b$ SISSA - International School for Advanced Studies, via Bonomea 265, I--34136 Trieste, Italy \\ \vspace{0.5cm}
$^c$Rechnergest\"utzte Physik der Werkstoffe, Institut f\"ur Baustoffe (IfB), \\ ETH Z\"urich, Stefano-Franscini-Platz 3,
CH - 8093 Z\"urich, Switzerland
\\ \vspace{0.5cm}
$^d$Centro de F\'{i}sica Te\'{o}rica e Computacional, Universidade de Lisboa, \\P--1749-016 Lisboa, Portugal
\\ \vspace{1.0cm}
\end{center}

\begin{abstract}
The leading asymptotic behaviour of the Humbert functions $\Phi_2$, $\Phi_3$, $\Xi_2$ of two variables is found, 
when the absolute values of the two independent variables become simultaneosly large. New integral representations of these functions are given. 
These are re-expressed as inverse Laplace transformations and the asymptotics is then found from a Tauberian theorem. 
Some integrals of the Humbert functions are also analysed. 
\end{abstract}

\vfill
\noindent
\textbf{keywords}: hypergeometric functions in two variables; Humbert function; asymptotics;
special functions; 
Tauberian theorem; many-body quantum systems

\vspace{1cm}
\noindent
MSC 2010 numbers: 
33C65, 33C70, 82C23, 33C20 \\~\\

\end{titlepage}

\setcounter{footnote}{0}

\section{Introduction}
Generalised hypergeometric functions, usually denoted by $_pF_q(z)$, 
and of which Gauss' hypergeometric function $_2F_1(z)$ is the most important special case, 
have been studied very thorougly and have found numerous applications in almost all fields of science, 
see e.g. \cite{Bailey35,Bateman53,Mathai73,Abra65,Slater66,Buchholz69,Seaborn91,Askey10,Mathai10} and refs. therein. A little more than a century old, 
hypergeometric functions of two variables \cite{Appell1880a,Appell1880b,Appell1882,Appell26,Humbert20a,Humbert20b} 
have also received a lot of 
scientific interest and recently, many new applications in many different fields of mathematics and
physics are being discovered see for example \cite{Fleischer03,Shpot07,Kniehl12}. 
It is often convenient to define these functions via double power series. 
Most of the mathematical studies of these functions are either focussed on the analysis of domains of convergence, or on relating special cases
to other known functions or else to derive functional relationships between different hypergeometric functions of two variables, see e.g. 
\cite{Appell26,Bateman53,Bateman54,Srivastava85,Srivastava89,Askey10,Choi11,Saxena11,Bry12,Bry14,Bry15,Bry17,Liu14,Brychkov17a,Brychkov17b}. 
Relatively little seems yet to be known on the asymptotic behaviour of such double series, in contrast to the classic study of
Wright \cite{Wright35,Wright40} on the asymptotics of $_pF_q(z)$ when $|z|\to \infty$. 
Here, we shall present results on the leading asymptotics of some hypergeometric functions 
when the absolute values of both variables become large simultaneously. 
The main tool to derive these are Eulerian and (inverse) Laplacian integral representations, and a Tauberian theorem \cite{Widder46,Feller71}.   
The results are stated as theorems in section~3, see eqs. (\ref{Phi2}-\ref{Phi3int},\ref{Phi2int}).

We shall consider the third Appell series $F_3$
\BEQ \label{1}
F_3(\alpha,\alpha',\beta,\beta';\gamma;x,y) = \sum_{m=0}^{\infty}\sum_{n=0}^{\infty} 
\frac{(\alpha)_m (\beta)_m (\alpha')_n (\beta')_n}{(\gamma)_{n+m}}\frac{x^m}{m!} \frac{y^n}{n!}
\EEQ
where $(\alpha)_m = \Gamma(\alpha+m)/\Gamma(\alpha)$ denotes the Pochhammer symbol for $-\alpha\not\in\mathbb{N}$. 
We shall also study the confluent forms (Humbert functions)
\begin{subequations} \label{2}
\begin{align}
\Xi_1(\alpha,\alpha',\beta;\gamma;x,y) &= \sum_{m=0}^{\infty}\sum_{n=0}^{\infty} 
\frac{(\alpha)_m (\beta)_m (\alpha')_n}{(\gamma)_{m+n}}\frac{x^m}{m!} \frac{y^n}{n!} \label{2Xi1} \\
\Xi_2(\alpha,\textcolor{black}{\beta};\gamma;x,y)       &= \sum_{m=0}^{\infty}\sum_{n=0}^{\infty} 
\frac{(\alpha)_m (\beta)_m}{(\gamma)_{m+n}}\frac{x^m}{m!} \frac{y^n}{n!}             \label{2Xi2} \\
\Phi_2(\beta,\beta';\gamma;x,y)        &= \sum_{m=0}^{\infty}\sum_{n=0}^{\infty} 
\frac{(\beta)_m (\beta')_n}{(\gamma)_{m+n}}\frac{x^m}{m!} \frac{y^n}{n!}             \label{2Phi2} \\
\Phi_3(\beta;\gamma;x,y)               &= \sum_{m=0}^{\infty}\sum_{n=0}^{\infty} 
\frac{(\beta)_m}{(\gamma)_{m+n}}\frac{x^m}{m!} \frac{y^n}{n!}                        \label{2Phi3}
\end{align}
\end{subequations} 
Throughout, we shall implicitly assume that the parameters $\alpha,\alpha',\beta,\beta',\gamma,\ldots$ are such that any singularity 
in the coefficients is avoided, 
without restating this explicitly. While the series $F_3$ converges for $\max( |x|, |y|)<1$, the series $\Xi_1$ and $\Xi_2$ converge for
$|x|<1$ and $|y|<\infty$ and $\Phi_2$ and $\Phi_3$ converge for $|x|<\infty$ and $|y|<\infty$ \cite{Srivastava85}. 
For reduction formul{\ae} to generalised hypergeometric functions of a single variable, see \cite{Brychkov17a,Brychkov17b}. 
We shall also be interested in the series 
\begin{subequations} \label{3}
\begin{align}
\Phi_2^{(i)}(\beta,\beta';\gamma,\lambda;x,y)  &:= \sum_{m=0}^{\infty}\sum_{n=0}^{\infty} 
\frac{(\beta)_m (\beta')_n}{(\gamma)_{m+n}}\frac{1}{m+n+\lambda}\frac{x^m}{m!} \frac{y^n}{n!}       
\label{3Phi2i} \\
\Phi_3^{(i)}(\beta;\gamma,\lambda;x,y)         &:= \sum_{m=0}^{\infty}\sum_{n=0}^{\infty} 
\frac{(\beta)_m}{(\gamma)_{m+n}}\frac{1}{m+n+\lambda}\frac{x^m}{m!} \frac{y^n}{n!}              
\label{3Phi3i}
\end{align}
\end{subequations}
which for $-\lambda\not\in\mathbb{N}$ converge for  $|x|<\infty$ and $|y|<\infty$. 
Clearly, for $\lambda=\gamma$, one has
\BEQ
\Phi_2^{(i)}(\beta,\beta';\gamma,\gamma;x,y) = \frac{1}{\gamma} \Phi_2(\beta,\beta';\gamma+1;x,y) \;\; , \;\; 
\Phi_3^{(i)}(\beta;\gamma,\gamma;x,y) = \frac{1}{\gamma} \Phi_3(\beta;\gamma+1;x,y)
\EEQ
and for $\lambda=1$ these series may be rewritten as Kamp\'e de F\'eriet series \cite{Srivastava85}
\begin{subequations} \label{5}
\begin{align}
\Phi_3^{(i)}(\beta;\gamma,1;x,y) &= \int_0^1 \!\D w\: \Phi_3( \beta;\gamma;xw, yw) 
= F_{2;0;0}^{1;1;0}\left(\left.\begin{array}{lll} (1); & (\beta); & -        \\ (\gamma,2); & -; & - \end{array}\right| x,y\right) \label{5Phi3i} \\[.5cm]
\Phi_2^{(i)}(\beta,\beta';\gamma,1;x,y) &= \int_0^1 \!\D w\: \Phi_2( \beta,\beta';\gamma;xw, yw) 
= F_{2;0;0}^{1;1;1}\left(\left.\begin{array}{lll} (1); & (\beta); & (\beta') \\ (\gamma,2); & -; & - \end{array}\right| x,y\right) \label{5Phi2i} 
\end{align}
\end{subequations}
such that $\Phi_2^{(i)}$, $\Phi_3^{(i)}$ might be called `integrated Humbert functions'. 
We are interested in situations where both $|x|$ and $|y|$ become large. 
We shall therefore substitute $x\mapsto -t x$ and $y \mapsto -t y$ and study the limit  
$t\to\infty$ where $x,y\in\mathbb{R}$ will be kept fixed and 
non-zero. The minus signs in the substitution rule are motivated from applications to quantum physics, 
see section~\ref{sec:application}.

In section~2 the integral and inverse Laplace representations of the Humbert functions and integrated Humbert functions are derived, 
and the integral representations are used to define the required analytic continuations. 
Furthermore, the inverse Laplace representations will be used in section~3 to derive the asymptotic forms. 
Section~4 briefly outlines an application to many-body quantum dynamics.

\section{Integral representations}

The starting point for the derivation of the asymptotics of the series (\ref{1}, \ref{2}) are the following Eulerian integral representations. 
Throughout, the parameters $\alpha,\beta,\gamma\ldots$ of the functions, as well as $x,y$, 
are assumed constants and such that all series and integrals considered exist.

\noindent \textcolor{black}{{\bf Lemma 1.} {\it 
Using the shorthand notations $\Phi_3 =\Phi_3(\beta;\gamma;-tx, -ty)$, 
$\Phi_2 =\Phi_2(\beta,\beta';\gamma;-tx,-ty)$, 
$\Xi_2  =\Xi_2(\alpha,\beta;\gamma;- tx,- ty)$, 
$\Xi_1  =\Xi_1(\alpha,\beta,\beta';\gamma;-tx,-ty)$ and  
$F_3 =F_3(\alpha,\alpha',\beta,\beta';\gamma;-tx,-ty)$, the functions defined in 
(\ref{1}, \ref{2}) have the integral representations}
\allowdisplaybreaks
\begin{subequations} \label{2.5}
\begin{align}
\Phi_3&= \frac{\Gamma(\gamma)\;t^{1-\gamma}}{\Gamma(\gamma-\vep)\Gamma(\vep)}
\int_0^{t} \!\D v\: v^{\gamma-\vep-1}\: {}_1F_1\left( \beta;\gamma-\vep;-xv\right)\; {}_0F_1\left(\vep; -y(t-v) \right) (t-v)^{\vep-1} \\[.25cm]
\Phi_2  &= \frac{\Gamma(\gamma)\;t^{1-\gamma}}{\Gamma(\gamma-\vep)\Gamma(\vep)}
\int_0^{t} \!\D v\: v^{\gamma-\vep-1}\: {}_1F_1\left( \beta;\gamma-\vep; -xv\right)\; 
{}_1F_1\left(\beta'; \vep; y(v-t) \right) (t-v)^{\vep-1}  \label{2.5Phi2} \\[.25cm]
\Xi_2 &= \frac{\Gamma(\gamma)\;t^{1-\gamma}}{\Gamma(\gamma-\vep)\Gamma(\vep)}
\int_0^{t} \!\D v\: v^{\gamma-\vep-1}\: {}_2F_1\left( \alpha,\beta;\gamma-\vep; -xv\right)\; 
{}_0F_1\left(\vep; -y(t-v) \right) (t-v)^{\vep-1} \label{2.5Xi2} \\[.25cm]
\Xi_1 &= 
\frac{\Gamma(\gamma)\;t^{1-\gamma}}{\Gamma(\gamma-\vep)\Gamma(\vep)}\int_0^{t} \!\D v\: v^{\gamma-\vep-1}\:
{}_2F_1\left( \alpha,\beta;\gamma-\vep; -xv\right)\; {}_1F_1\left(\beta'; \vep; -y(t-v) \right)(t-v)^{\vep-1} \label{2.5Xi1} \\[.25cm]
\nonumber
F_3&= \frac{\Gamma(\gamma)\;t^{1-\gamma}}{\Gamma(\gamma-\vep)\Gamma(\vep)} 
\int_0^{t} \!\D v\: v^{\gamma-\vep-1}\:
{}_2F_1\left(\alpha,\beta;\gamma-\vep;- xv\right)\;\times\\
&\hspace{6cm}\times{}_2F_1\left(\alpha',\beta'; \vep; -y(t-v) \right) (t-v)^{\vep-1} \label{2.5F3}
\end{align}
\end{subequations}
{\it and where $\vep$ is a fixed constant, which satisfies $0<\vep<\gamma$.}}

The integral representations (\ref{2.5}) were already given in \cite{Bry15} and (\ref{2.5F3}) in \cite[(9.4.16)]{Srivastava85}. 
We begin by repeating them since they are the crucial starting point of our analysis.

\noindent
{\bf Proof:} \textcolor{black}{We illustrate the technique for the example $\Phi_2$. 
The double series (\ref{2Phi2}) is decoupled
by using the decomposition $m+n+\gamma= (m+\gamma-\vep)+(n+\vep)$ and the identity \cite[(6.2.1)]{Abra65} involving the Euler Beta function
\BEA
\frac{1}{\Gamma(m+n+\gamma)} 
= \frac{B(n+\vep,m+\gamma-\vep)}{\Gamma(n+\vep) \Gamma(m+\gamma-\vep)}  
= \frac{\int_0^1\D u\; (1-u)^{n+\vep-1}u^{m+\gamma-\vep-1}}{\Gamma(n+\vep) \Gamma(m+\gamma-\vep)}\label{2.2}
\EEA
Inserting this into the definition (\ref{2Phi2}) gives, because the series are absolutely convergent 
\begin{align} \nonumber
\Phi_2 &= \frac{\Gamma(\gamma)}{\Gamma(\gamma-\vep) \Gamma(\vep)}\int_0^1\D u \; 
\sum_{m=0}^{\infty}\frac{(\beta)_m}{(\gamma-\vep)_m}\frac{(-t x u)^m}{m!}
\sum_{n=0}^{\infty}\frac{(\beta')_n}{(\vep)_n}\frac{(-t y (1-u))^n}{n!}
\frac{u^{\gamma-\vep-1}}{(1-u)^{1-\vep}}\\[.25cm]
&= \frac{\Gamma(\gamma)}{\Gamma(\gamma-\vep)\Gamma(\vep)} \int_0^{1} \!\D u \; (1-u)^{\vep-1} u^{\gamma-\vep-1}\:
{}_1F_1\left( \beta;\gamma-\vep;-txu\right) {}_1F_1\left(\beta'; \vep; -ty(1-u) \right)
\nonumber
\end{align}
The assertion (\ref{2.5Phi2}) follows by rescaling $v=t u$. The other identities eqs.~(\ref{2.5}) are derived similarly.} \hfill {\bf q.e.d.}

\noindent
{\bf Comment.} Recall the following definitions of some further double hypergeometric series \cite{Srivastava85}
\begin{subequations} \label{2.3}
\begin{align}
F_2(a,b,b';c,c';x,y) &= \sum_{m=0}^{\infty} \sum_{n=0}^{\infty} \frac{(a)_{m+n} (b)_m (b')_n}{(c)_m (c')_n} \frac{x^m}{m!} \frac{y^n}{n!} \\[.25cm]
\Psi_1(a,b;c,c';x,y) &= \sum_{m=0}^{\infty} \sum_{n=0}^{\infty} \frac{(a)_{m+n} (b)_m}{(c)_m (c')_n} \frac{x^m}{m!} \frac{y^n}{n!} \\[.25cm]
\Psi_2(a;c,c';x,y)   &= \sum_{m=0}^{\infty} \sum_{n=0}^{\infty} \frac{(a)_{m+n}}{(c)_m (c')_n} \frac{x^m}{m!} \frac{y^n}{n!} 
\end{align}
\end{subequations}
By using $\Gamma(z) = \int_0^{\infty} \!\D u\: u^{z-1} e^{-u}$, one may derive in a way similar to Lemma 1 the identities
\begin{subequations} \label{2.4}
\begin{align}
F_2(a,b,b';c,c';x,y) &= \frac{1}{\Gamma(a)}\int_0^{\infty} \!\D u\: e^{-u}\: u^{a-1}\: {}_1F_1(b;c;x u) {}_1F_1(b';c';y u) \\[.25cm]
\Psi_1(a,b;c,c';x,y) &= \frac{1}{\Gamma(a)}\int_0^{\infty} \!\D u\: e^{-u}\: u^{a-1}\: {}_1F_1(b;c;x u) {}_0F_1(c';y u) \\[.25cm]
\Psi_2(a;c,c';x,y)   &= \frac{1}{\Gamma(a)}\int_0^{\infty} \!\D u\: e^{-u}\: u^{a-1}\: {}_0F_1(c;x u) {}_0F_1(c';y u)
\end{align}
\end{subequations}
see also \cite[(9.4.29)]{Srivastava85}, \cite[(27)]{Bry14}. 
However, there is no known direct way to render these as convolutions, which will become our main tool to analyse
the $t\to\infty$ asymptotics of the functions in Lemma 1. A different route is suggested by \cite[eq. (31)]{Bry14}. \\

The integral representations (\ref{2.5Xi2},\ref{2.5Xi1},\ref{2.5F3}) can be used to extend the definition of the functions $\Xi_2, \Xi_1, F_3$ 
(unecessary for $\Phi_2,\Phi_3$). 
This is based on the Eulerian integral representation, for $\gamma>\beta>0$ \cite[eq. (15.3.1)]{Abra65}
\BEQ
{}_2F_1\left(\alpha,\beta;\gamma;-x\right) = \frac{\Gamma(\gamma)}{\Gamma(\beta)\Gamma(\gamma-\beta)}
\int_0^1 \!\D u\: u^{\beta-1} (1-u)^{\gamma-\beta-1} (1+ux)^{-\alpha}
\EEQ
which defines the analytic continuation of the function ${}_2F_1\left(\alpha,\beta;\gamma;-x\right)$ in the domain $|\arg x|<\pi$, which has a cut for
$-1>x>-\infty$ \cite{Bateman53}. \\

\noindent
{\bf Definition:} {\it The integral representations (\ref{2.5}) define the (principal branch of) the functions $\Phi_3 =\Phi_3(\beta;\gamma;-tx, -ty)$, 
$\Phi_2 =\Phi_2(\beta,\beta';\gamma;-tx,-ty)$, $\Xi_2  =\Xi_2(\alpha,\beta;\gamma;- tx,- ty)$, 
$\Xi_1  =\Xi_1(\alpha,\beta,$ $\beta'; \gamma;-tx,-ty)$ and  $F_3 =F_3(\alpha,\alpha',\beta,\beta';\gamma;-tx,-ty)$ in the regions $|\arg xt| <\pi$ and
$|\arg yt| <\pi$. They are the analytical
continuations of the series (\ref{1},\ref{2}), beyond their respective domain of convergence.} 

Indeed, from (\ref{2.5}), the functions $\Phi_2, \Phi_3$ are defined for all $|x|<\infty$, $|y|<\infty$, 
while $\Xi_1$, $\Xi_2$ have a cut for $-1>xt>-\infty$ and $F_3$ has cuts for $-1>xt>-\infty$ and $-1>yt>-\infty$. All results which follow
will implicitly use these analytic continuations. \\

In consequence, if we use $\mathscr{F}$ as a generic symbol for any of the functions in (\ref{2.5}), one can recast (\ref{2.5}) as follows (in the sense of 
the analytic continuation) 
\begin{align}\nonumber
\mathscr{F}(t) &= \frac{\Gamma(\gamma)}{\Gamma(\gamma-\vep)\Gamma(\vep)} t^{1-\gamma} \int_0^t \!\D v\: \mathscr{F}_1(v) \mathscr{F}_2(t-v) \\
&= \frac{\Gamma(\gamma)}{\Gamma(\gamma-\vep)\Gamma(\vep)} t^{1-\gamma} 
\mathscr{L}^{-1}\left( \overline{\mathscr{F}}_1(p)\: \overline{\mathscr{F}}_2(p)\right)(t)\label{2.6}
\end{align}
where $\overline{f}(p) = \mathscr{L}\left( f(v) \right)(p) = \int_0^{\infty}\!\D v\: e^{-pv} f(v)$ denotes the Laplace transform. 
The functions $\mathscr{F}_{1,2}(v)$ are readily read off from eqs.~(\ref{2.5}) and are listed in  the following table.  

\begin{center}\begin{tabular}{l|ll}
$\mathscr{F}(t)$ & \multicolumn{1}{c}{$\mathscr{F}_1(v)$} &  \multicolumn{1}{c}{$\mathscr{F}_2(v)$} \\[0.10truecm] \hline
$\Phi_3$ & $v^{\gamma-\vep-1}\: {}_1F_1\left( \beta;\gamma-\vep; -xv\right)$ & $v^{\vep-1}\: {}_0F_1\left(\vep; -yv \right)$ \\[0.25truecm]
$\Phi_2$ & $v^{\gamma-\vep-1}\: {}_1F_1\left( \beta;\gamma-\vep; -xv\right)$ & $v^{\vep-1}\: {}_1F_1\left(\beta'; \vep; -yv \right)$ \\[0.25truecm]
$\Xi_2$  & $v^{\gamma-\vep-1}\: {}_2F_1\left( \alpha,\beta;\gamma-\vep; -xv\right)$ & $v^{\vep-1}\: {}_0F_1\left(\vep; -yv \right)$ \\[0.25truecm] 
$\Xi_1$  & $v^{\gamma-\vep-1}\: {}_2F_1\left( \alpha,\beta;\gamma-\vep; -xv\right)$ & $v^{\vep-1}\: {}_1F_1\left(\beta'; \vep; -yv \right)$ \\[0.25truecm] 
$F_3$    & $v^{\gamma-\vep-1}\: {}_2F_1\left( \alpha,\beta;\gamma-\vep; -xv\right)$ & $v^{\vep-1}\: {}_2F_1\left(\alpha',\beta'; \vep; -yv\right)$ \\[0.25truecm]
\hline 
\end{tabular}\end{center}

\noindent
Next, we require the following list of Laplace transforms, taken from \cite[(3.35.1.3,3.37.1.2,3.38.1.1)]{Prud4}, combined with \cite[(13.1.10,13.1.33)]{Abra65}
\begin{subequations} \label{lapF}
\begin{align}
\mathscr{L}\left( v^{a-1}\: {}_0F_1\left( a; -yv\right)\right)(p) &= \Gamma(a) p^{-a} e^{-y/p} \label{lapFa}\\[.25cm]
\mathscr{L}\left( v^{b-1}\: {}_1F_1\left( a;b; -yv\right)\right)(p) &= \Gamma(b) p^{a-b} (p+y)^{-a} \label{lapFb}\\[.1cm]
\mathscr{L}\left( v^{c-1}\: {}_2F_1\left( a,b;c; -yv\right)\right)(p) &= \Gamma(c) p^{a-c} y^{-a} U\left(a;1+a-b;\frac{p}{y}\right) \label{lapFc}
\end{align}
\end{subequations}
where $U$ denotes the Tricomi function \cite{Abra65}. Combining these with the integral forms (\ref{2.6}) gives 

\noindent
{\bf Lemma 2.} {\it The Laplace transforms of the analytically continued functions in \textcolor{black}{Lemma 1} are given by the following table, where $U$ denotes the Tricomi function.}

\noindent
\begin{center}\begin{tabular}{ll}
function $\mathscr{F}(t)/\left(\Gamma(\gamma) t^{1-\gamma}\right)$ & Laplace transform $\overline{\mathscr{F}}(p)$ \\[0.10truecm] \hline
$\Phi_3(\beta;\gamma;-xt, -yt)$         & $p^{\beta-\gamma} (p+x)^{-\beta} e^{-y/p}$ \\[0.25truecm]
$\Phi_2(\beta,\beta';\gamma; -xt, -yt)$ & $p^{\beta+\beta'-\gamma} (p+x)^{-\beta} (p+y)^{-\beta'}$ \\[0.25truecm]
$\Xi_2(\alpha,\beta;\gamma;-xt, -yt)$   & $x^{-\alpha} p^{\alpha-\gamma} U(\alpha;1+\alpha-\beta;p/x) e^{-y/p}$ 
\\[0.25truecm]
$\Xi_1(\alpha,\beta,\beta';\gamma;-xt, -yt)$ & $ x^{-\alpha} p^{\alpha+\beta'-\gamma} (p+y)^{-\beta'} 
U(\alpha;1+\alpha-\beta;p/x)$  \\[0.25truecm]
$F_3(\alpha,\alpha',\beta,\beta';\gamma;-xt, -yt)$ & $x^{-\alpha} y^{-\alpha'} p^{\alpha+\alpha'-\gamma} 
                                                               U(\alpha;1+\alpha-\beta;\frac{p}{x})U(\alpha';1+\alpha'-\beta';\frac{p}{y})$  \\[0.25truecm]
\hline
\end{tabular}
\end{center}
The entries for $\Phi_2$ and $\Phi_3$ are contained in \cite{Bateman54}.\\

\noindent
{\bf Corollary 1.} \textcolor{black}{Applying again eq.~(\ref{2.2}), the formal Kamp\'e de F\'eriet series}
{\small
\begin{align}\nonumber
&\hspace{-.3cm}
 F_{1;q;q'}^{0;p;p'}\hspace{-.15cm}\left(\hspace{-.25cm}\left.\begin{array}{lll} -;\  (\alpha_p);\  (\alpha'_{p'}) \\[.25cm]\ \gamma;\  (\beta_q);\   (\beta'_{q'})\end{array}
\hspace{-.2cm}\right| -tx, -ty\hspace{-.05cm}\right)\hspace{-.1cm}=\hspace{-.25cm}\sum_{m,n=0}^{\infty} \frac{(\alpha_1)_m \cdots (\alpha_p)_m}{(\beta_1)_m\cdots (\beta_q)_m} 
\frac{(\alpha'_1)_n \cdots (\alpha_{p'})_n}{(\beta'_1)_n\cdots (\beta'_{q'})_n} \frac{(-1)^{m+n}}{(\gamma)_{m+n}} 
\frac{(tx)^m}{m!} \frac{(ty)^n}{n!}
\nonumber \\[.5cm]\nonumber
=&\ \textcolor{black}{\frac{\Gamma(\gamma)}{\Gamma(\gamma-\vep)\Gamma(\vep)} 
\int_0^{1}\hspace{-.05cm} \frac{ \!\D u\: u^{\gamma-\vep-1}}{(1-u)^{1-\vep}}\: 
{}_pF_{q+1}\left(\begin{array}{l} \alpha_1, \ldots, \alpha_p \\ \beta_1,\ldots,\beta_q,\gamma-\vep \end{array}; {-txu}\right)} \times \label{2.8a} \\ 
& \hspace{7cm}\times
{}_{p'}F_{q'+1}\left(\begin{array}{l} \alpha'_1, \ldots, \alpha'_{p'} \\ \beta'_1,\ldots,\beta'_{q'},\vep \end{array}; {-ty(1-u)}\right) 
 \\[.5cm]
=& \ \Gamma(\gamma) t^{1-\gamma} \mathscr{L}^{-1}\left( s^{-\gamma}\: {}_pF_q\left( (\alpha_p); (\beta_q); -\frac{x}{s}\right) 
{}_{p'}F_{q'}\left( (\alpha'_{p'}); (\beta'_{q'}); -\frac{y}{s}\right)\right)(t) 
\label{2.8b}
\end{align}
} 
\noindent \hspace{-0.15truecm}contains all functions treated here explicitly as special cases, if (\ref{2.8a}) can be used as above 
to define an analytic continuation, in  the domain $|\arg xt|<\pi$ and $|\arg yt|<\pi$. 
For the derivation of (\ref{2.8b}), we used the identity \cite[(3.38.1.1)]{Prud4}
\BEQ \label{2.9}
\mathscr{L}\left( v^{\mu-1}\: {}_pF_{q+1}\left( (a_p); (b_q), \mu; - \omega v \right)\right)(s) = \Gamma(\mu) s^{-\mu}\:
{}_pF_q\left( (a_p); (b_q); - \frac{\omega}{s} \right)
\EEQ
For $q=q'=0$ and $p=p'=2$, eq.~(\ref{2.8a}) reduces to $F_3$, or the Lauricella function $F_B^{(2)}$ in two variables. 
Furthermore, for $p=p'=q=q'=0$, one has an addition theorem
\begin{align}\nonumber
{}_0F_1(\gamma;x+y)&= \sum_{m,n=0}^{\infty} \frac{1}{(\gamma)_{m+n}}\frac{x^m}{m!}\frac{y^n}{n!} \\
&= \textcolor{black}{\frac{\Gamma(\gamma)}{\Gamma(\gamma-\vep)\Gamma(\vep)}\int_0^{1}\!\D u\:\frac{u^{\gamma-\vep-1}}{(1-u)^{1-\vep}}\; 
{}_0F_1\left(\gamma-\vep;{xu}\right)\:{}_0F_1\left(\vep;{y(1-u)}\right)}
\end{align}

We now turn to the variants $\Phi_3^{(i)}$ and $\Phi_2^{(i)}$ defined in eq.~(\ref{3}). Since $\Phi_2^{(i)}$ is symmetric under the
permutation $(x,\beta) \leftrightarrow (y,\beta')$, we can set $x\geq y$ without restriction of the generality. 

\noindent
{\bf Lemma 3.} \textcolor{black}{{\it With the shorthands $\Phi_3^{(i)} = \Phi_3^{(i)}(\beta;\gamma,1;-tx, -ty)$, 
$\Phi_2^{(i)} = \Phi_2^{(i)}(\beta,\beta';\gamma,1;-tx, -ty)$ with $x>y$ and $\Phi_2^{(i,s)} := 
\Phi_2^{(i)}(\beta,\beta';\gamma,1;-tx, -tx)$, the following integral representations  of the integrated Humbert functions (\ref{3}) hold true}} 
\begin{subequations} \label{lapPhi}
\begin{align}
&\Phi_3^{(i)} = \Gamma(\gamma) t^{1-\gamma} \frac{e^{y/x}}{x}\left(\frac{y}{x}\right)^{\beta-1} 
\mathscr{L}^{-1}\bigg( p^{1-\gamma} \bigg[ \Gamma\left(1-\beta,\frac{y}{x}\right)- \Gamma\left(1-\beta,\frac{y}{x}+\frac{y}{p}\right)\bigg] \bigg)(t)
\label{lapPhi3i_int}\\[.35cm]
&\Phi_2^{(i)} = \frac{\Gamma(\gamma) t^{1-\gamma} x^{\beta'-1}}{(1-\beta) (x-y)^{-\beta'}}\;
\mathscr{L}^{-1}\left(- p^{1-\gamma}\:{}_2F_1\left(1-\beta,\beta';2-\beta;-\frac{y}{(x-y)}\right)\right.
\nonumber \\
& \hspace{4truecm}\left. + p^{\beta+\gamma}(p+x)^{1-\beta}\: {}_2F_1\left(1-\beta,\beta';2-\beta;-\frac{(p+x)y}{p(x-y)}\right) \right)(t)
\label{lapPhi2i_int} \\[.35cm]
&\Phi_2^{(i,s)} 
= \frac{\Gamma(\gamma) t^{1-\gamma}}{(1-\beta-\beta') x}\; \mathscr{L}^{-1}\left(
p^{\beta+\beta'-\gamma}(p+x)^{1-\beta-\beta'} - p^{1-\gamma}\right)(t)
\label{lapPhi2i_int_symm} 
\end{align}
\end{subequations}
{\it where $\Gamma(a,x)$ is the incomplete Gamma-function.}

\noindent
{\bf Proof.} Starting from (\ref{3Phi3i}), the extra denominator is turned into an auxiliary integral
\BD
\Phi_3^{(i)} = \int_0^{\infty} \!\D v\: \sum_{m,n=0}^{\infty} e^{-v(m+n+1)} \frac{(\beta)_m}{(\gamma)_{m+n}} \frac{(-tx)^m}{m!}\frac{(-ty)^n}{n!}
\ED
and the decoupling of the two series proceeds via (\ref{2.2}), as in  the proof of Lemma 1. A last change of variables $w=e^{-v}$ 
and using also (\ref{lapF}) leads to 
\BD
\Phi_3^{(i)} = \Gamma(\gamma) t^{1-\gamma} \mathscr{L}^{-1}\bigg( p^{\beta-\gamma} 
\underbrace{\int_0^1 \!\D w\: (p+xw)^{-\beta} e^{-yw/p}}_{=:{\cal M}} \bigg)(t)
\ED
The integral $\cal M$ is found as follows, reducing it to incomplete Gamma functions \cite{Abra65}
\begin{align}
{\cal M} &= \frac{1}{x} \int_p^{x+p} \!\D a\: a^{-\beta}\: \exp\left[-\frac{y}{x}\frac{a-p}{p}\right]
\:=\: \frac{e^{y/x}}{x} \left(\frac{px}{y}\right)^{1-\beta} \int_{y/x}^{y/x+y/p} \!\D b\: b^{-\beta} e^{-b}
\nonumber \\
&= \frac{e^{y/x}}{x} \left(\frac{px}{y}\right)^{1-\beta} 
\left[ \Gamma\left(1-\beta,\frac{y}{x}\right) - \Gamma\left(1-\beta,\frac{y}{x}+\frac{y}{p}\right)\right]
\nonumber
\end{align}
and inserting into $\Phi_3^{(i)}$ gives the assertion (\ref{lapPhi3i_int}). 

Turning to $\Phi_2^{(i)}$, the procedure to go from (\ref{3Phi2i}) to an integral representation follows the same lines as before. 
Changing variables as before and re-using (\ref{lapF}), we find 
\BD
\Phi_2^{(i)} = \Gamma(\gamma) t^{1-\gamma} \mathscr{L}^{-1}\bigg( p^{\beta+\beta'-\gamma} 
\underbrace{\int_0^1 \!\D w\: (p+xw)^{-\beta} (p+yw)^{-\beta'}}_{=: {\cal N}} \bigg) (t)
\ED
which is still symmetric under the simultaneous exchanges $(x,\beta) \leftrightarrow (y,\beta')$, as it should be. 
To evaluate this, recall the following identity on the incomplete Beta function \cite[(6.6.8,15.3.4)]{Abra65}
\BD
I(a,b,\xi) := \int_0^{\xi} \!\D u \: \frac{u^{a-1}}{(1+u)^b} = \frac{\xi^a}{a} \; {}_2F_1(a,b;1+a;-\xi)
\ED
Then we can evaluate the integral $\cal N$, now using $x>y$
\begin{align}
{\cal N} &= x^{-\beta} y^{-\beta'} \int_0^1 \!\D w\: \left( w+\frac{p}{x}\right)^{-\beta} \left( w+\frac{p}{y}\right)^{-\beta'} 
\nonumber \\[.25cm]
&= x^{-\beta} y^{-\beta'} \left(\frac{p(x-y)}{xy}\right)^{1-\beta-\beta'}
\int_{y/(x-y)}^{(p+x)y/(p(x-y))} \!\D b\: b^{-\beta} (1+b)^{-\beta'}
\nonumber \\[.25cm]
&= x^{-\beta} y^{-\beta'} \left(\frac{p(x-y)}{xy}\right)^{1-\beta-\beta'}
\left[ I\left(1-\beta,\beta', \frac{(p+x)y}{p(x-y)}\right) - I\left(1-\beta,\beta', -\frac{y}{(x-y)}\right)\right]
\nonumber \\[.25cm]
&=  \left(\frac{p(x-y)}{xy}\right)^{1-\beta-\beta'} \frac{x^{-\beta} y^{-\beta'}}{1-\beta} 
\left[ \left(\frac{(p+x)y}{p(x-y)}\right)^{1-\beta}\:{}_2F_1\left(1-\beta,\beta';2-\beta;-\frac{(p+x)y}{p(x-y)}\right) \right.
\nonumber \\[.25cm]
&  \left.\hspace{5.5cm}- \left(\frac{y}{x-y}\right)^{1-\beta}\:{}_2F_1\left(1-\beta,\beta';2-\beta;-\frac{y}{x-y}\right)\right]
\nonumber
\end{align}
and inserting this into the above expression for $\Phi_2^{(i)}$ gives the assertion (\ref{lapPhi2i_int}). 
Finally, in the symmetric case $x=y$ we have 
\BD
\Phi_2^{(i,s)} = \Gamma(\gamma) t^{1-\gamma} \mathscr{L}^{-1}\left( p^{\beta+\beta'-\gamma} 
\int_0^1 \!\D w\: (p+xw)^{-\beta-\beta'}  \right) (t)
\ED
and straightforward integration gives the assertion (\ref{lapPhi2i_int_symm}). \hfill { \bf q.e.d.} 

\noindent
{\bf Corollary 2.} {\it For $x>0$, and $\lambda>0$, $\mu>0$, one has the identity}
\begin{equation}
 \label{Corrollar2}
\int_0^1  \frac{\!\D w\: w^{\lambda-1}}{(1-w)^{1-\mu}}\: \Phi_2(\beta,\beta';\gamma; -xw, -xw) = 
\frac{\Gamma(\lambda)\Gamma(\mu)}{\Gamma(\lambda+\mu)}\; {}_2F_2\left(\beta+\beta',\lambda;\mu+\lambda,\gamma;-x\right)
\end{equation}
{\bf Proof.} The lines of calculation follow the proof of $\Phi_2^{(i,s)}$ in Lemma 3. Consider 
\begin{align}
\int_0^1 \!\D w\  & w^{\lambda-1} (1-w)^{\mu-1} \Phi_2(\beta,\beta';\gamma; -xtw, -xtw)  \nonumber \\[.25cm]
&= \Gamma(\gamma) t^{1-\gamma} \mathscr{L}^{-1}\left( p^{\beta+\beta'-\gamma} 
\int_0^1 \!\D w\: w^{\lambda-1} (1-w)^{\mu-1} (p+xw)^{-\beta-\beta'} \right)(t) \nonumber \\[.25cm]
&= \frac{\Gamma(\gamma)\Gamma(\lambda)\Gamma(\mu)}{\Gamma(\lambda+\mu)}\, t^{1-\gamma}
\mathscr{L}^{-1}\left( p^{-\gamma}\: {}_2F_1\left(\beta+\beta',\lambda;\mu+\lambda;-\frac{x}{p}\right) \right)(t) \nonumber \\[.25cm]
&= \frac{\Gamma(\lambda)\Gamma(\mu)}{\Gamma(\lambda+\mu)}\: {}_2F_2\left(\beta+\beta',\lambda;\mu+\lambda,\gamma;-xt\right)
\nonumber
\end{align}
where in the third line, the integral representation \cite[(15.3.1)]{Abra65} of $_2F_1$ was used and in the forth line, 
\cite[(3.35.1.10)]{Prud5}, or else (\ref{2.9}), was applied. Set $t=1$. \hfill { \bf q.e.d.} 

Similar, but inequivalent, integral formulae involving $\Phi_2$ 
are stated in \cite[eqs. (3.18,3.19)]{Bry12}.

\section{Asymptotic expansions}

We shall use a Tauberian theorem for the asymptotic evaluations: {\it the behaviour of a function $f(t)$ for $t\to\infty$ is related
to the one of its Laplace transform $\overline{f}(p)$ for $p\to 0$} \cite{Widder46}, \cite[ch. XIII]{Feller71}. Therefore, it is sufficient to analyse the
behaviour of the representations as inverse Laplace transformations from Lemmas 2 and 3 in section~2 for $p\to 0$, before inverting. 

\noindent
{\bf Theorem 1.} {\it The Humbert function $\Phi_2=\Phi_2(\beta,\beta';\gamma;-tx,-ty)$ has the following leading asymptotic behaviour for $t\to\infty$, with
$x,y\ne 0$ being kept fixed}
\BEQ \label{Phi2}
\Phi_2 \simeq \left\{
\begin{array}{ll} 
\frac{\Gamma(\gamma)}{\Gamma(\gamma-\beta-\beta')} \left( tx\right)^{-\beta} \left( ty\right)^{-\beta'} & \mbox{\rm ~~;~ for $x>0$, $y>0$} \\[0.5truecm] 
\frac{\Gamma(\gamma)}{\Gamma(\beta')}\: e^{-yt} \left( t(|y|+x)\right)^{-\beta} \left( t|y|\right)^{\beta+\beta'-\gamma} & \mbox{\rm ~~;~ for $x>0$, $y<0$} \\[0.5truecm] 
\frac{\Gamma(\gamma)}{\Gamma(\beta)}\: e^{-xt} \left( t(y+|x|)\right)^{-\beta'} \left( t|x|\right)^{\beta+\beta'-\gamma} & \mbox{\rm ~~;~ for $x<0$, $y>0$} \\[0.5truecm] 
\frac{\Gamma(\gamma)}{\Gamma(\beta)}\: e^{-|x|t} \left( t|x|\right)^{\beta+\beta'-\gamma} \left( t|x-y|\right)^{-\beta'} & \mbox{\rm ~~;~ for $x<y<0$} \\[0.5truecm] 
\frac{\Gamma(\gamma)}{\Gamma(\beta')}\: e^{-|y|t} \left( t|y|\right)^{\beta+\beta'-\gamma} \left( t|y-x|\right)^{-\beta'} & \mbox{\rm ~~;~ for $y<x<0$} \\[0.5truecm] 
\frac{\Gamma(\gamma)}{\Gamma(\beta+\beta')}\: e^{-|x|t} \left( t|x|\right)^{\beta+\beta'-\gamma} & \mbox{\rm ~~;~ for $y=x<0$}  
\end{array} \right.
\EEQ
{\it and neither $\gamma$, $\beta$, $\beta'$, $\beta+\beta'$ nor $\gamma-\beta-\beta'$ are non-positive integers.}

Only the signs of $\beta,\beta'$ and of $\beta+\beta'-\gamma$ will influence the qualitative behaviour of the leading asymptotic terms, for $t\to\infty$. 

\noindent
{\bf Proof.} The starting point is the representation of $\Phi_2$ as an inverse Laplace transformation from 
\textcolor{black}{Lemma~2}. For $x>0$ and $y>0$, the leading term
for $p\to 0$ is found to be $\Phi_2\simeq \Gamma(\gamma) t^{1-\gamma} \mathscr{L}^{-1}\left( p^{\beta+\beta'-\gamma} x^{-\beta} y^{-\beta'}\right)(t)$ 
and direct evaluation \cite[(2.1.1.1)]{Prud5} gives the assertion. Next, for $x>0$ and $y<0$, one first makes the shift $p=q-y$
\textcolor{black}{which produces an exponential contribution, according to the {\it shift theorem} }
\BEQ\nonumber
\mathscr{L}^{-1}\left(\bar{f} (p-\omega)\right)(t) = \e^{\omega t } f(t)
\EEQ
Second, one takes the leading term for $q\to 0$. Then 
\begin{align}
\Phi_2 &=  \Gamma(\gamma) t^{1-\gamma} e^{-|y|t} 
\mathscr{L}^{-1}\left( (q+|y|)^{\beta+\beta'-\gamma} (q+|y|+x)^{-\beta} q^{-\beta'}\right)(t) \nonumber\\[.15cm]
&\stackrel{q\to 0}{\simeq} \Gamma(\gamma) t^{1-\gamma} e^{-|y|t} 
\mathscr{L}^{-1}\left( |y|^{\beta+\beta'-\gamma} (|y|+x)^{-\beta} q^{-\beta'}\right)(t) \nonumber
\end{align}
and direct evaluation gives the assertion. For $x<0$ and $y>0$ one merely has to permute $(x,\beta)\leftrightarrow (y,\beta')$. 
Finally, for $x<0$ and $y<0$
\BD
\Phi_2 = \Gamma(\gamma) t^{1-\gamma} \mathscr{L}^{-1}\left( p^{\beta+\beta'-\gamma} (p-|x|)^{-\beta} (p-|y|)^{-\beta'}\right)(t)
\ED
If $x<y<0$, or $|x|>|y|$, one makes the shift $q=p-|x|$ and the stated result follows as before. If $y<x<0$, one merely permutes 
$(x,\beta)\leftrightarrow (y,\beta')$. For $x=y<0$, the shift $q=p-|x|$ and expansion in $q$ to lowest order gives the stated result. \hfill {\bf q.e.d.} 

\noindent
{\bf Theorem 2.} {\it The Humbert function $\Phi_3=\Phi_3(\beta;\gamma;-tx,-ty)$ has the following asymptotic behaviour for $t\to\infty$, with
$x,y\ne 0$ being kept fixed}
\BEQ \label{Phi3}
\Phi_3 \simeq \left\{
\begin{array}{ll} 
\Gamma(\gamma) (tx)^{-\beta} (ty)^{(1+\beta-\gamma)/2} J_{\gamma-\beta-1}(2\sqrt{yt\,}\,) & \mbox{\rm ~~;~ for $x>0$, $y>0$} \\[0.5truecm] 
\Gamma(\gamma) (tx)^{-\beta} (t|y|)^{(1+\beta-\gamma)/2} I_{\gamma-\beta-1}(2\sqrt{|y|t\,}\,) & \mbox{\rm ~~;~ for $x>0$, $y<0$} \\[0.5truecm]
\frac{\Gamma(\gamma)}{\Gamma(\beta)} \left( t|x|\right)^{\beta-\gamma} e^{-y/|x| - |x|t} & \mbox{\rm ~~;~ for $x<0$}
\end{array} \right.
\EEQ
{\it where $J_{\nu}$ is a Bessel function and $I_{\nu}$ the corresponding modified Bessel function and neither 
$\gamma$ nor $\beta$ are non-positive integers.} 

The qualitative behaviour of the leading asymptotic term is only influenced by the signs of $\beta$ and $\beta-\gamma$. 

\noindent
{\bf Proof.} Use the representation of $\Phi_3$ as an inverse Laplace transformation in \textcolor{black}{Lemma~2}. For $x>0$, simply retain the lowest order
in $p\to 0$ and use (\ref{lapFa}). Expressing the hypergeometric function ${}_0F_1$ as a Bessel or a modified Bessel function, respectively, gives
the assertion for $y>0$ and $y<0$. For $x<0$, make the shift $q=p-|x|$ such that
\begin{align}\nonumber
\Phi_3 &=  \Gamma(\gamma) t^{1-\gamma}\: e^{-|x|t} \mathscr{L}^{-1}\left( (q+|x|)^{\beta-\gamma} q^{-\beta} e^{-y/(q+|x|)}\right)(t)\\[.2cm] \nonumber
&\stackrel{q\to 0}{\simeq} \Gamma(\gamma) t^{1-\gamma}\: e^{-|x|t} \mathscr{L}^{-1}\left( |x|^{\beta-\gamma} q^{-\beta} e^{-y/|x|}\right)(t)
\end{align}
and re-use $\Gamma(\beta)\mathscr{L}^{-1}\left( q^{-\beta}\right)(t)=t^{\beta-1}$ \cite[(2.1.1.1)]{Prud5}. \hfill { \bf q.e.d.}

\noindent
{\bf Theorem 3.} {\it The Humbert function $\Xi_2=\Xi_2(\alpha,\beta;\gamma;-tx,-ty)$ has the following asymptotic behaviour for $t\to\infty$, with
$x,y\ne 0$ and $x>0$ being kept fixed}
\BEQ \label{Xi2}
\Xi_2 \simeq \left\{
\begin{array}{ll} 
\frac{\Gamma(\alpha)\Gamma(\alpha-\beta)}{\Gamma(\alpha)} \left( tx\right)^{-\beta}\left( ty\right)^{-(\gamma-\beta-1)/2} 
J_{\gamma-\beta-1}(2\sqrt{yt\,}\,) & \mbox{\rm ;~ $\forall$ $y>0$, $\alpha>\beta$} \\[0.5truecm]
\frac{\Gamma(\alpha)\Gamma(\alpha-\beta)}{\Gamma(\alpha)} \left( tx\right)^{-\beta}\left( t|y|\right)^{-(\gamma-\beta-1)/2} 
I_{\gamma-\beta-1}(2\sqrt{|y|t\,}\,) & \mbox{\rm ;~ $\forall$ $y<0$, $\alpha>\beta$} \\[0.5truecm]
\frac{\Gamma(\gamma)}{\Gamma(\alpha)} \left( tx\right)^{-\alpha}\left( ty\right)^{-(\gamma-\alpha-1)/2} 
\left[ \frac{\pi}{2} Y_{\gamma-\alpha-1}(2\sqrt{yt\,}\,)\right. & \\[0.07truecm] 
~~\left. +J_{\gamma-\alpha-1}(2\sqrt{|y|t\,}\,) 
\left[ \demi \ln(tx) + \ln(x/y) -\psi(\alpha) -2C_E\right]\right]  & \mbox{\rm ;~ $\forall$ $y>0$, $\alpha=\beta$} \\[0.5truecm]
\frac{\Gamma(\gamma)}{\Gamma(\alpha)} \left( tx\right)^{-\alpha}\left( t|y|\right)^{-(\gamma-\alpha-1)/2} 
I_{\gamma-\alpha-1}(2\sqrt{|y|t\,}\,) & \\[0.07truecm] 
~~\times \left[ \demi \ln(tx) + \ln(x/|y|) -\psi(\alpha) -2C_E\right]  & \mbox{\rm ;~ $\forall$ $y<0$, $\alpha=\beta$} \\[0.5truecm]
\frac{\Gamma(\alpha)\Gamma(\beta-\alpha)}{\Gamma(\beta)} \left( tx\right)^{-\alpha}\left( ty\right)^{-(\gamma-\alpha-1)/2} 
J_{\gamma-\alpha-1}(2\sqrt{yt\,}\,) & \mbox{\rm ;~ $\forall$ $y>0$, $\alpha<\beta$} \\[0.5truecm]
\frac{\Gamma(\alpha)\Gamma(\beta-\alpha)}{\Gamma(\beta)} \left( tx\right)^{-\alpha}\left( t|y|\right)^{-(\gamma-\alpha-1)/2} 
I_{\gamma-\alpha-1}(2\sqrt{|y|t\,}\,) & \mbox{\rm ;~ $\forall$  $y<0$, $\alpha<\beta$} 
\end{array} \right.
\EEQ
{\it where $J_{\nu}$ and $Y_{\nu}$ are the Bessel and Neuman functions, respectively, $I_{\nu}$ is a modified Bessel function, 
$\psi(x)$ is the digamma function and $C_E\simeq 0.5772\ldots$ is Euler's constant \cite{Abra65}. For $x<0$, the function $\Xi_2$ has a cut.} 

\noindent
{\bf Proof.} In order to apply the inverse Laplace representation of \textcolor{black}{Lemma~2}, the small-$p$ expansion
\BD
U\left(\alpha;1+\alpha-\beta;\frac{p}{x}\right) \simeq \left\{ \begin{array}{ll}
\frac{\Gamma(\alpha-\beta)}{\Gamma(\alpha)} \left( \frac{p}{x}\right)^{\beta-\alpha} & \mbox{\rm ~~;~ for $\alpha>\beta$} \\[0.25truecm]
-\frac{1}{\Gamma(\alpha)}\left[ \ln\frac{p}{x} + \psi(\alpha) + 2C_E \right]         & \mbox{\rm ~~;~ for $\alpha=\beta$} \\[0.25truecm]
\frac{\Gamma(\beta-\alpha)}{\Gamma(\beta)}                                           & \mbox{\rm ~~;~ for $\alpha<\beta$} 
\end{array} \right. 
\ED
according to \cite[(13.5.6-13.5.12)]{Abra65} is required, for $\alpha-\beta\not\in -\mathbb{N}$. For $x>0$ and $\alpha>\beta$, to lowest order
in $p\to 0$, this gives 
$\Xi_2 \simeq \frac{\Gamma(\gamma)\Gamma(\alpha-\beta)}{\Gamma(\alpha)} t^{1-\gamma} x^{-\beta} \mathscr{L}^{-1}\left( p^{-(\gamma-\beta)} e^{y/p}\right)(t)$
and using (\ref{lapFa}) gives the assertion. For $x>0$ and $\alpha<\beta$ the result follows from the symmetry in $\alpha$ and $\beta$. 
For $\alpha=\beta$ and $y>0$, expansion to lowest order in $p\to 0$ gives
\begin{align}\nonumber
\Xi_2 &\simeq -\frac{\Gamma(\gamma)}{\Gamma(\alpha)} \frac{t^{1-\gamma}}{ x^{\alpha}} \bigg[ 
\mathscr{L}^{-1}\left( \left(\psi(\alpha)+2C_E-\ln x\right) p^{-(\gamma-\alpha)} e^{-y/p}\right)(t)\\ 
&\hspace{7cm}+ \mathscr{L}^{-1}\left( p^{-(\gamma-\alpha)} \ln p\; e^{-y/p}\right)(t) \bigg]\nonumber \\[.5cm]
&= -\frac{\Gamma(\gamma)}{\Gamma(\alpha)} \frac{t^{1-\gamma}}{ x^{\alpha}} \left[ 
\left(\psi(\alpha)+2C_E-\ln x -\demi \ln\frac{t}{y}\right) \left(\frac{t}{y}\right)^{\demi(\gamma-\alpha-1)} J_{\gamma-\alpha-1}(2\sqrt{yt\,}\,) \right.
\nonumber \\
&  \left. \hspace{7cm}- \left(\frac{t}{y}\right)^{\demi(\gamma-\alpha-1)} 
\left. \frac{\partial J_{\nu-1}(2\sqrt{yt\,}\,)}{\partial \nu}\right|_{\nu=\gamma-\alpha} \right]
\nonumber
\end{align}
re-using (\ref{lapFa}) and \cite[(2.5.7.3)]{Prud5}. For $z\to\infty$, one has asymptotically 
$\frac{\partial J_{\nu}(z)}{\partial \nu}\simeq \frac{\pi}{2} Y_{\nu}(z)$ \cite[(9.25,9.26)]{Abra65}. 
Collecting terms leads to the stated result. For $y<0$ and $\alpha=\beta$ one has analogously
\begin{align}\nonumber
\Xi_2 &\simeq -\frac{\Gamma(\gamma)}{\Gamma(\alpha)}\frac{ t^{1-\gamma}}{ x^{\alpha}} \bigg[ 
\mathscr{L}^{-1}\left( \left(\psi(\alpha)+2C_E-\ln x\right) p^{-(\gamma-\alpha)} e^{|y|/p}\right)(t) \\
& \hspace{8cm} +\mathscr{L}^{-1}\left( p^{-(\gamma-\alpha)} \ln p\; e^{|y|/p}\right)(t) \bigg]\nonumber \\[.5cm]
&= -\frac{\Gamma(\gamma)}{\Gamma(\alpha)} \frac{t^{1-\gamma}}{ x^{\alpha} }\left[ 
\left(\psi(\alpha)+2C_E-\ln x -\demi \ln\frac{t}{y}\right) 
\left(\frac{t}{y}\right)^{\demi(\gamma-\alpha-1)} I_{\gamma-\alpha-1}(2\sqrt{yt\,}\,) \right.
\nonumber \\
&  \left. \hspace{6.75cm}- \left(\frac{t}{y}\right)^{\demi(\gamma-\alpha-1)} 
\left.\frac{\partial I_{\nu-1}(2\sqrt{yt\,}\,)}{\partial \nu}\right|_{\nu=\gamma-\alpha} \right]
\nonumber
\end{align}
and from the asymptotic form \cite[(9.7.1)]{Abra65} for $I_{\nu}(z)$ for $z\to\infty$, we see that 
$\frac{\partial I_{\nu}(z)}{\partial \nu}\simeq -\frac{\nu}{z} I_{\nu}(z)$ merely gives a sub-leading correction. Collecting
terms we complete the list of assertions if $x>0$. For $x<0$, the inverse Laplace representation in \textcolor{black}{Lemma 2} has a cut. 
\hfill {\bf q.e.d.} 

\noindent
{\bf Theorem 4.} {\it The integrated Humbert function $\Phi_3^{(i)}=\Phi_3^{(i)}(\beta;\gamma,1;-tx,-ty)$ 
has the following leading asymptotic behaviour for $t\to\infty$, with $x,y\ne 0$ and $x>0$ being kept fixed}
\BEQ
\Phi_3^{(i)} \simeq \left\{
\begin{array}{ll} 
\frac{\gamma-1}{xt}\left[ \Gamma(1-\beta) \left(y/x)\right)^{\beta-1} - \frac{1}{1-\beta}\:{}_1F_1(1;2-\beta;y/x)\right] 
& \mbox{\rm ;~ $\forall$  $y>0$, $\beta+\gamma>\frac{3}{2}$} \\[0.5truecm]
\frac{\Gamma(\gamma)}{\sqrt{\pi\,}\,} \left( yt \right)^{-\demi(\beta+\gamma+\demi)} \left(\frac{y}{x}\right)^{\beta} 
\cos\left( 2\sqrt{yt\,}\, +\frac{\pi}{2}\left(\beta-\gamma-\demi\right)\right) 
& \mbox{\rm ;~ $\forall$  $y>0$, $\beta+\gamma<\frac{3}{2}$} \label{Phi3int} \\[0.5truecm]
\frac{\Gamma(\gamma)}{2\sqrt{\pi\,}\,} \left( |y|t \right)^{-\demi(\beta-\gamma-\demi)} \left( xt\right)^{-\beta} \exp\left(2\sqrt{|y|t\,}\,\right) 
& \mbox{\rm ;~ $\forall$  $y<0$} \\[0.11truecm]
\end{array} \right.
\EEQ
{\it where neither $\gamma$ nor $1-\beta$ are non-positive integers.} 

\noindent
{\bf Proof.} Begin with the integral representation (\ref{lapPhi3i_int}) of \textcolor{black}{Lemma 3}. The leading term for $p\to 0$ is found from the
asymptotic identity \cite[(6.5.30)]{Abra65} for $x\to\infty$
\BD
\Gamma(a,x+y) - \Gamma(a,x) \simeq -e^{-x} x^{a-1} \left( 1 - e^{-y}\right)
\ED
In order to invert $\mathscr{L}$, we also need the identities (\ref{lapFa}) and \cite[(3.10.2.2)]{Prud5} 
\BD
\mathscr{L}^{-1}\left( p^{-\mu}\Gamma\left(\nu,\frac{a}{p}\right)\right)(t) = \frac{\Gamma(\nu)}{\Gamma(\mu)}t^{\mu-1} -
\frac{a^{\nu}t^{\mu+\nu-1}}{\nu\Gamma(\mu+\nu)}\:{}_1F_2\left(\nu;\nu+1,\mu+\nu;-at\right)
\ED
Then, for $p\to 0$ (here, $x>0$ is assumed) 
\begin{align}\nonumber
\Phi_3^{(i)} &\simeq \Gamma(\gamma)t^{1-\gamma} 
\frac{e^{y/x}}{x} \left(\frac{y}{x}\right)^{\beta-1} 
\hspace{-.2cm}\mathscr{L}^{-1}\bigg( \frac{\Gamma(1-\beta,\frac{y}{x})}{p^{\gamma-1}} \\ &\hspace{5cm}- \frac{\Gamma(1-\beta,\frac{y}{p})}{p^{\gamma-1}} 
+ \frac{e^{-y/p}}{p^{\gamma-1}} \left(\frac{p}{y}\right)^{\beta} \left( 1 - e^{-y/x}\right) \bigg)(t) 
\nonumber \\[.5cm]
&=\Gamma(\gamma)t^{1-\gamma} \frac{e^{y/x}}{x} \left(\frac{y}{x}\right)^{\beta-1} \left[ 
\frac{\Gamma(1-\beta,y/x)-\Gamma(1-\beta)}{\Gamma(\gamma-1)} \right. \nonumber \\
& \hspace{.5cm}  \left.+
\frac{(yt)^{\beta-1}{}_1F_2(1-\beta;2-\beta,\gamma-\beta;-yt)}{(1-\beta)\Gamma(\gamma-\beta)}\:+ 
\frac{(1-e^{-y/x}){}_0F_1(\gamma-\beta-2;-yt)}{\Gamma(\gamma-\beta-1)(yt)^{\beta}}\: \right]\nonumber
\end{align}
Further evaluation is simplified by the identity, taken from \cite[(6.5.3,6.5.12)]{Abra65}
\BD
\left( \frac{y}{x}\right)^{-(1-\beta)} \left[ \Gamma(1-\beta) - \Gamma\left(1-\beta,\frac{y}{x}\right)\right] = \frac{1}{1-\beta}\:
{}_1F_1\left(1-\beta;2-\beta;-\frac{y}{x}\right)
\ED
and this gives
\begin{align}
\Phi_3^{(i)} &\simeq \Gamma(\gamma) \frac{e^{y/x}}{xt} \left[ 
\frac{{}_1F_1(1-\beta;2-\beta;-y/x)}{(\beta-1)\Gamma(\gamma-1)} \right. \nonumber \\
&  \left.+
\frac{{}_1F_2(1-\beta;2-\beta,\gamma-\beta;-yt)}{(1-\beta)\Gamma(\gamma-\beta) (xt)^{\beta-1}} +
\frac{1}{(xt)^{\beta}}\frac{x}{y}\frac{1-e^{-y/x}}{\Gamma(\gamma-\beta-1)}\:{}_0F_1(\gamma-\beta-2;-yt)\right] ~~~
\label{3.5}
\end{align}
We can now distinguish the two cases $y>0$ and $y<0$. For $y>0$, recall the asymptotic identity \cite[(07.22.06.0011.01)]{Wolfram}
\begin{align}
_1F_2(a_1;b_1,b_2;-y) &\stackrel{y\to\infty}{\simeq} \frac{\Gamma(b_1)\Gamma(b_2)}{\sqrt{\pi}\,\Gamma(a_1)}\: y^{\eta/2}
\cos\left(\frac{\pi}{2}\eta+2\sqrt{y\,}\right) \left( 1 + {\rm O}(y^{-\frac{1}{2}})\right) \nonumber \\
& \hspace{3cm} + \frac{\Gamma(b_1)\Gamma(b_2)}{\Gamma(b_1-a_1)\Gamma(b_2-a_1)}\: y^{-a_1} \left( 1 + {\rm O}(y^{-1})\right) \nonumber
\end{align}
with $\eta = (a_1-b_1-b_2+\demi)$. Also, the function $_0F_1$ can be expressed in terms of Bessel functions $J_{\nu}$ \cite{Abra65}. 
Inserting into (\ref{3.5}) the above expansion and using the asymptotics of $J_\nu$ \cite{Abra65} leads to
\begin{align} 
\Phi_3^{(i)} &\simeq
\frac{\gamma-1}{xt} \left[ \Gamma(1-\beta) \left(\frac{y}{x}\right)^{\beta-1} - \frac{1}{1-\beta}\:{}_1F_1\left(1;2-\beta;\frac{y}{x}\right) \right]
\nonumber \\
& \hspace{3cm} + \frac{\Gamma(\gamma)}{\sqrt{\pi}\,}\left(\frac{y}{x}\right)^{\beta} \left( yt\right)^{-\demi(\beta+\gamma+\demi)}
\cos\left( 2\sqrt{yt\,}\,+\frac{\pi}{2}\left(\beta-\gamma-\demi\right)\right)
\nonumber
\end{align}
Herein, the first line dominates for $\beta+\gamma>\frac{3}{2}$ and the second line for $\beta+\gamma<\frac{3}{2}$. 
This is the first part of the assertion. 
For $y=-|y|<0$, recall the asymptotic form \cite[(07.22.06.0005.01)]{Wolfram}
\BD
_1F_2(a_1;b_1,b_2;y) \stackrel{y\to\infty}{\simeq} \frac{\Gamma(b_1)\Gamma(b_2)}{2\sqrt{\pi}\,\Gamma(a_1)}\: y^{\demi(a_1-b_1-b_2+\demi)}\, e^{2 \sqrt{y}}
\ED
and now express $_0F_1$ in terms of a modified Bessel function $I_{\nu}$ \cite{Abra65}. Insertion into (\ref{3.5}) 
and using the known asymptotic behaviour leads to
\BD
\Phi_3^{(i)} \simeq \frac{\Gamma(\gamma)}{2\pi^{1/2}} \frac{(|y|t)^{-\demi(\beta-\gamma-\demi)}}{(xt)^{\beta}}\, e^{2\sqrt{|y|t\,}}
+\frac{\gamma-1}{\beta-1}\frac{1}{xt}\: {}_1F_1\left(1;2-\beta;-\frac{|y|}{x}\right)
\ED
Clearly, the second term is always sub-dominant. This completes the proof.  \hfill { \bf q.e.d.}

\noindent
{\bf Theorem 5.} {\it The integrated Humbert function $\Phi_2^{(i)}=\Phi_2^{(i)}(\beta,\beta';\gamma,1;-tx,-ty)$ 
has the following leading asymptotic behaviour for $t\to\infty$, with $x>y>0$ being kept fixed}
\begin{align}
\Phi_2^{(i)} &\simeq \frac{\Gamma(\gamma)}{(1-\beta)\Gamma(\gamma-\beta)}\frac{(xt)^{\beta'-\beta}}{\left( (x-y)t\right)^{\beta}}\:
{}_2F_2\left(1-\beta,\beta';2-\beta,\gamma-\beta;-\frac{xy}{x-y}t\right) \nonumber \\
& \hspace{3cm} +\frac{\gamma-1}{\beta-1}\frac{1}{xt}\left(\frac{x}{x-y}\right)^{\beta'}\:
{}_2F_1\left(1-\beta,\beta';2-\beta;-\frac{y}{x-y}\right) 
\label{Phi2int}
\end{align}
{\it  Herein, none of $\gamma$, $\gamma-\beta$ or $1-\beta$ is a non-positive integer. For $y>x>0$, one permutes $(x,\beta)$ and $(y,\beta')$.} 

\noindent
{\bf Proof.} Begin with the integral representation (\ref{lapPhi2i_int}) of \textcolor{black}{Lemma 3}. For $p\to 0$, this simplifies into
\begin{align}
\Phi_2^{(i)} &\stackrel{p\to 0}{\simeq} \frac{\Gamma(\gamma)}{1-\beta}\frac{t^{1-\gamma}}{(x-y)^{\beta'} x^{1-\beta'}} \;
\mathscr{L}^{-1}\left( \frac{x^{1-\beta}}{p^{\gamma-\beta}}\:{}_2F_1\left(1-\beta,\beta';2-\beta;-\frac{xy}{x-y}\frac{1}{p}\right) \right.
\nonumber \\
&  \left. \hspace{3cm}-p^{\gamma-1}\:{}_2F_1\left(1-\beta,\beta';2-\beta;-\frac{y}{x-y}\right)\right)(t)
\nonumber
\end{align}
We need the identity \cite[(3.35.1.10)]{Prud5}
\BD
\mathscr{L}^{-1}\left( p^{\nu}\:{}_2F_1\left(a,b;c;-\frac{\omega}{p}\right)\right)(t) = \frac{t^{\nu-1}}{\Gamma(\nu)}\:{}_2F_2(a,b;c,\nu;-\omega t)
\ED
Then straightforward algebra leads to the assertion. For $y>x>0$, it is enough to exchange $\beta\leftrightarrow\beta'$ and $x\leftrightarrow y$.  
\hfill {\bf q.e.d.}

\textcolor{black}{The symmetric case $x=y>0$, hence $\Phi_2^{(i,s)}$, is a special case of the corrollary~2, eq.~(\ref{Corrollar2}).} 
More explicit asymptotics of $_2F_2$ can be found in \cite{Wright40,Wolfram,Volkmer14}. 

These expressions derived in this section can be checked numerically. 
However, the convergence towards the given asymptotics is in  general quite slow. 

Finally, it is now straightforward to obtain the asymptotics of the special Kamp\'e de F\'eriet series (\ref{2.8b}), by using  
the known asymptotics of the generalised hypergeometric functions $_pF_q(z)$ \cite{Wright40}. 

\section{An example from physics}
\label{sec:application}
The quantum spherical model \cite{Henkel84,Vojta96,Oliv06} is a simple exactly solvable model of quantum phase transitions, in $d$ spatial dimensions, 
with a non-trivial quantum critical behaviour at zero temperature 
(that is, the model cannot be described by a simple mean-field approximation, at least for $1<d<3$), see e.g. 
\cite{Sachdev11,Dutta15}. 
Its main formal characteristic is the `spherical constraint'. If the coherent and dissipative quantum dynamics of the model is formulated 
in terms of a Lindblad equation, it can be shown that the canonical quantum commutation relations are maintained, in spite of the dissipation, 
at least on average. 
\textcolor{black}{While already the dynamical single-body problem is of physical {\it and} mathematical interest \cite{Wald16}, the 
full $N$-body problem makes explicit use of the here derived asymptotic descriptions.}
If the system \textcolor{black}{is quenched from a highly symmetric initial state deep into} the ordered phase, 
in  the $N\to\infty$ limit the spherical constraint takes the form $\mathscr{I}_1 + \mathscr{I}_2 =1$, where \cite{Wald17}
\BEA
\mathscr{I}_1 &=& \int_{\cal B}\frac{\D\vec{k}}{(2\pi)^d}\: e^{-\gamma(Z+t\omega_{\vec{k}})} \:=\: 
e^{-\gamma Z} \left( e^{-2\gamma t} I_0(2\gamma t)\right)^d \:\stackrel{t\to\infty}\simeq\: e^{-\gamma Z} \left(4\pi\gamma t\right)^{-d/2} \label{4.1} \\
\mathscr{I}_2 &=& \demi \int_{\cal B}\frac{\D\vec{k}}{(2\pi)^d}\: 
\left( 1 - \frac{Cgt}{Z+t\omega_{\vec{k}}}\right)\left( 1 - \cos 2 \vartheta_{\vec{k}} \right) e^{-\gamma(Z+t\omega_{\vec{k}})}
\label{4.2}
\EEA
where $Z=Z(t)$ is the integrated spherical Lagrange multiplier whose long-time behaviour for $t\to\infty$ is sought. \textcolor{black}{Furthermore, 
$\omega_{\vec{k}} = 2(d-\cos k_1 -\ldots - \cos k_d)$ is the lattice dispersion relation on a $d$-dimensional hypercubic lattice with nearest-neighbour
interactions, ${\cal B}=[-\pi,\pi]^d$ is the $d$-dimensional Brillouin zone, 
and $\vartheta_{\vec{k}}=\sqrt{ gt\: (Z+t\omega_{\vec{k}})}$. Finally, the constant $g$ is the 
quantum coupling, $\gamma$ describes the dissipative coupling to an external reservoir and $C$ characterises the initial (disordered) quantum state.}
The \textcolor{black}{integral} $\mathscr{I}_2$ can be evaluated by expanding 
the cosine and integrating termwise. Then the spherical constraint can be rewritten in the form \cite{Wald17}
\BEQ \label{4.3}
e^{\gamma Z} (4\pi\gamma t)^{d/2} = \demi \Phi_3\left(\frac{d}{2};\frac{3}{2};-g Z t, -\frac{g}{\gamma} t\right) 
+ C g^2 t^2 \int_0^1 \!\D w\: \Phi_3\left(\frac{d}{2}; \frac{3}{2}; -\frac{g}{\gamma} t w, -g Z t w\right)
\EEQ
For the physically interesting long-time behaviour of $Z=Z(t)$ for $t\to\infty$,  the asymptotics of the Humbert function $\Phi_3$ and of the
integrated Humbert function $\Phi_3^{(i)}$, as studied in this work, are required.  In contrast to the original formulation in eqs.~(\ref{4.1},\ref{4.2}), 
the reformulation in eq.~(\ref{4.3}) contains the spatial dimension $d$ merely as a parameter. This allows to discuss also the model's behaviour
at non-integer dimensions $d\in\mathbb{R}$, which often provides useful physical insight. 

The final long-time behaviour obtained from (\ref{4.3}) turns out to depend subtly on the dimension $d$. For $d\geq 2$, there is 
a single solution with $Z(t)=-|Z(t)|<0$. Then, for $t\to\infty$ it follows that $|Z(t)|\sim t^{-1}\ln^2 t$ for $d>2$ and $|Z(t)|\sim t^{-1}$ for
$d=2$. In both cases, this is quite different from the form $|Z(t)|\sim \ln t$ 
obtained for a classical, non-coherent dynamics (limit $g\to 0$) \cite{Wald17}. 
~\\~\\

\section*{Acknowledgements}
We are grateful to the  Group `Rechnergest\"utzte Physik der Werkstoffe' at ETH Z\"urich, Switzerland, where this work was done, 
for their warm hospitality. 
SW thanks UFA-DFH for financial support through grant CT-42-14-II. 

\end{document}